\documentclass[12pt]{JHEP3}
\pdfoutput=1

\usepackage{graphicx,subfigure}
\usepackage{amsmath}

\newcommand{\bea}{\begin{eqnarray}}
\newcommand{\eea}{\end{eqnarray}}
\newcommand{\la}{\langle}
\newcommand{\ra}{\rangle}

\renewcommand{\a}{\alpha}
\renewcommand{\b}{\beta}
\renewcommand{\c}{\chi}
\renewcommand{\d}{\delta}
\renewcommand{\k}{\kappa}
\renewcommand{\l}{\lambda}
\newcommand{\m}{\mu}
\newcommand{\n}{\nu}
\renewcommand{\r}{\rho}
\newcommand{\s}{\sigma}

\newcommand{\rh}{r_H}

\title{\begin{center} 
Holographic Superconductors and Higher Curvature Corrections
\end{center}}
\author{
Massimo Siani
\\~
\\
Dipartimento di Fisica, Universit\`a di Milano Bicocca\\
and \\
INFN, Sezione di Milano-Bicocca,\\ 
piazza della Scienza 3, I-20126 Milano, Italy\\
~~\\
\email{massimo.siani@mib.infn.it}
}

\abstract{ We study a fully backreacted holographic model of a four-dimensional superconductor by including a higher curvature interaction in the bulk action. We study how the critical temperature and the field theory condensate vary in this model and conclude that positive higher curvature couplings make the condensation harder. We also compute the conductivity, finding significant deviations from the conjectured universal frequency gap to critical temperature ratio.
}

\preprint{}

\begin{document}

\section{Introduction}
The gauge/gravity correspondence \cite{Maldacena:1997re} has proven to be a powerful tool for examining the properties of strongly coupled field theories. In the recent literature it has been applied to the study of condensed matter systems providing a good qualitative description of their quantum phase transition to a superconducting phase \cite{Hartnoll:2009sz,Herzog:2009xv}.

The superconducting phase of a field theory is characterised by a charged operator which condenses for sufficiently low temperatures, breaking a local symmetry. According to the AdS/CFT correspondence, its vacuum expectation value is mapped to a charged field which assumes a nontrivial profile in the dual theory in the bulk. To take into account the fact that the field theory is at finite temperature, the gravitational background is an asymptotically AdS black hole which is unstable when its Hawking temperature is lower than a critical temperature. It has been shown that there is indeed the possibility that a scalar field spontaneously breaks a $U(1)$ symmetry near the AdS black hole horizon in the bulk \cite{Gubser:2005ih,Gubser:2008px}. The AdS/CFT correspondence relates the general asymptotic falloff of the scalar field near the boundary to the vacuum expectation value of the dual field theory operator. Based on this observation, a simple model of a $(2+1)$-dimensional superconductor was constructed \cite{Hartnoll:2008vx,Hartnoll:2008kx}. It is remarkable that despite the fact that the gravitational model was not derived from a string theory background, it captures the essential physics of the holographic superconductors. Many aspects of holographic superconductivity are now well understood also in the $(3+1)$-dimensional case, such as their behavior in the presence of an external magnetic field, different scalar masses, potentials and backgrounds; for a (partial) list see \cite{Horowitz:2008bn,Gubser:2009cg,Horowitz:2009ij,Konoplya:2009hv,Gubser:2008wv,Peeters:2009sr,Maeda:2009wv,Nakano:2008xc,Albash:2008eh,Ge:2010aa,Takahashi:2009dz,Kanno:2010pq,Brihaye:2010mr}.

The possibility of obtaining string- (and M-)derived models of holographic superconductivity has been explored in \cite{Gubser:2009qm,Gauntlett:2009dn}. However, due to our lack of understanding of string theory backgrounds, the class of dual field theories that one can study with holography is in some way restricted. A way to overcome this problem is to consider higher curvature (or derivative) interactions in an effective gravitational theory. From the field theory point of view the higher curvature interactions introduce new couplings amongst the field theory operators, thus broadening the class of field theories one can holographically study. The higher curvature terms are treated in a perturbative framework, because they are seen as a $\a^\prime$ expansion of the full string theory action. When this expansion breaks down, one has to consider again the full string theory model.

If we restrict our attention to five dimensional gravity, the first step to take into account stringy corrections to the gravitational action is to include the Gauss-Bonnet term. It is believed to be the first, ${\cal O}(\a^\prime)$, correction to low energy gravity. Models of holographic superconductivity have been studied in this setting \cite{Gregory:2009fj,Pan:2009xa,Pan:2010at,Cai:2010cv,Barclay:2010up}. In the Einstein-Gauss-Bonnet gravity the metric equations of motion remain second order in (time) derivatives and planar AdS black holes can be analytically studied. A natural way to extend this model would be to include the cubic interactions of Lovelock gravity. However, this term only affects the equations of motion in seven or higher dimensions, and its effects are only visible in six-dimensional dual field theory. Inspired by the form of the Gauss-Bonnet equations of motion, recently a cubic theory of gravity has been constructed and AdS black holes solutions analytically found \cite{Oliva:2010eb,Myers:2010ru}. This model is obtained by adding a curvature-cubed interaction to the Einstein-Gauss-Bonnet action. Due to the fact that this curvature-cubed term has not a topological origin like the Lovelock ones, the linearized equations of motion for a graviton perturbation are fourth order in derivatives. Nevertheless, when the background metric is taken to be an AdS spacetime, they reduce to second order and this makes the holographic theory under control. Moreover, the full equations of motion for a static spherically symmetric metric in five dimensions reduce to a simple algebraic equation. For these reasons the model has been called quasi-topological gravity. This model has been considered for holographic studies in \cite{Myers:2010jv,Amsel:2010aj,Myers:2010xs,Sinha:2010pm}.

In this paper, we use the quasi-topological model to build a model of holographic superconductivity. Holographic superconductors in quasi-topological gravity have been studied in \cite{Kuang:2010jc} in the probe limit. Here, we also take into account the effects of the backreaction of the fields on the metric. Our main motivation relies in the recently noted features of Gauss-Bonnet holographic superconductors. In particular, it has been noted \cite{Horowitz:2008bn} that the real part of the conductivity of a holographic superconductor has a frequency gap which is proportional to the critical temperature of the system. The authors found that the frequency gap to critical temperature ratio assumes a constant value for every field theory with a gravitational Einstein dual without any reference to the field theory dimensionality or to the bulk scalar mass, so they claimed that this value should be universal. However, when the higher curvature terms are included, this relation ceases to hold. This is true both in the probe limit \cite{Gregory:2009fj,Pan:2009xa,Pan:2010at,Cai:2010cv} and in the backreacted case \cite{Barclay:2010up} for the Gauss-Bonnet model, and in the probe limit of the quasi-topological gravity \cite{Kuang:2010jc}. Here we show that it is also the case in the fully backreacted model of quasi-topological gravity. From our and other results, it seems that the frequency gap to the critical temperature ratio increases both with the Gauss-Bonnet and the quasi-topological couplings. The opposite behavior has been noted for the shear viscosity to the entropy density ratio: in that case, the $1/4\pi$ bound is violated by the higher curvature terms in such a way that increasing the gravitational parameters lowers the ratio.

The rest of the paper is organized as follows. In section \ref{thebulktheory} we review the quasi-topological holographic superconductor, derive the set of equations of motion and give some of the known analytical solutions. In section \ref{numerics} we present our numerical results for the condensate value and the critical temperature of the dual field theory. In section \ref{sec:conductivity} we analyze the conductivity of the holographic superconductor. We conclude in section \ref{conclusions}.

\section{The bulk theory} \label{thebulktheory}
We consider the following five-dimensional bulk action of a cubic theory of gravity \cite{Oliva:2010eb,Myers:2010ru} coupled to an abelian gauge field $A_\m$ and a charged massive scalar field $\phi$
\bea
S = && S_g + S_m \label{bulkaction} \\
S_g = && \frac{1}{2 \k^2}\int d^5 x \sqrt{-g} \left( R + \frac{12}{L^2} + \frac{\a L^2}{2} \c_4 + 3 \b L^4 {\cal L}_5 \right) \nonumber \\
S_m = && \int d^5 x \sqrt{-g} \left( -\frac{1}{4} F_{\m\n} F^{\m\n} - \left| \nabla_\m \phi - i q A_\m \phi \right|^2 - m^2 | \phi|^2 \right) \nonumber
\eea
where $\kappa^2=8 \pi G_5$,
\bea
\c_4 &&= R_{\m\n\r\s} R^{\m\n\r\s} - 4 R_{\m\n} R^{\m\n} + R^2 \\
{\cal L}_5 &&= -\frac{7}{6}R_{\ \ \r\s}^{\m\n}R_{\ \ \n\l}^{\r\d}R_{\ \ \m\d}^{\s\l}%
-R_{\m\n}^{\ \ \r\s}R_{\r\s}^{\ \ \n\l}R^{\m}_{\ \l}-\frac{1}{2}R_{\m\n}^{\ \ \r\s}%
R_{\ \r}^{\m}R_{\ \s}^{\n} \nonumber \\
&& ~~~~~~~~~~~+\frac{1}{3}R_{\ \n}^{\m}R_{\ \r}^{\n}R_{\ \m}^{\r}-\frac
{1}{2}RR_{\ \n}^{\m}R_{\ \m}^{\n}+\frac{1}{12}R^{3}
\eea
are the Gauss-Bonnet term and the curvature-cubed term we are considering\footnote{the curvature-cubed term is not unique: one can always modify it by adding the six-dimensional Euler density which does not affect the equations of motion in five dimensions \cite{Myers:2010ru}.}, respectively, and $\a$ and $\b$ are the Gauss-Bonnet coupling and the quasi-topological coupling, respectively. For simplicity, we take $\a$ and $\b$ to be real and positive. The equations of motion which follow from the action (\ref{bulkaction}) are \cite{Oliva:2010eb}
\bea
G_{\m\n} -\frac{6}{L^2} g_{\m\n} + \frac{\a L^2}{2} GB_{\mu\nu} + 3 \b L^4 E_{\mu\nu} = \k^2 T_{\m\n}
\label{gravEOM}
\eea
where $G_{\m\n}$ is the Einstein tensor,
\bea
GB_{\m\n} = 2RR_{\m\n} - 4 R_{\m\r} R_{\ \n}^{\r} - 4 R_{\ \r}^{\d} R_{\ \m\d\n}^{\r} + 2 R_{\m\r\d\gamma} R_{\n}^{\ \r\d\gamma} - \frac{1}{2} g_{\m\n} \c_4
\eea
\bea
E_{\m \n }  &&  = -\frac{7}{6} \left[  3R_{\m hd}^{\ \ \ g}R_{\n }^{\ prd}%
R_{pgr}^{\ \ \ h}-3\nabla _{p}\nabla _{q}(R_{\ g\ h}^{p\ q}R_{\m \ \n }%
^{\ g\ h}-R_{\ h\n g}^{p}R_{\m }^{\ gqh}) \right] \nonumber \\
&& -\Big[ R_{\m c\n d}R^{cspq}R_{pqs}^{\ \ \ d}-R_{\m }^{\ qcd}R_{cd\n }%
^{\ \ \ h}R_{qh}+R_{\n }^{\ dqc}R_{\m dc}^{\ \ \ h}R_{qh}  \nonumber \\
&& ~~~-\nabla _{p}\nabla_{q} \Big( R_{\m h}R_{\n }^{\ phq}+R_{\m h}R_{\n }^{\ qhp}+R_{\n h}R_{\m }^{\ phq} +R_{\ h}^{q}R_{\m \ \n }^{\ h\ p}+R_{\ h}^{p}R_{\m \ \n }^{\ q\ h}  \nonumber \\
&& ~~~ \left. \left. +\frac{1}{2}g^{pq}R_{\m }^{\ hcd}R_{\n }^{\ hcd}+\frac{1}{2}g_{\m \n } R^{prcd} R_{\ rcd}^{q} - g_{\m }^{\ p}R_{\n }^{\ rcd}R_{\ rcd}^{q} \right) \right] \nonumber \\
&&  -\frac{1}{2}\Big[ R_{\m c}R_{\n }^{\ fcd}R_{fd}+2R_{\m c\n d}R^{cfdg}R_{fg} \nonumber \\
&& ~~~ +\nabla _{p}\nabla _{q}(R_{\m \n }R^{pq}-R_{\m }^{\ p}R_{\n }^{\ q}+g^{pq}R_{\m c\n d}%
R^{cd}+g_{\m \n }R^{pcqd}R_{cd} -2g_{\m }^{\ p}R_{\ c\n d}^{q}R^{cd}) \Big] \nonumber \\
&& +\frac{1}{3}\left[ 3R_{\m c\n d}R^{ec}R_{e}^{\ d}+\nabla _{p}\nabla _{q}(\frac
{3}{2}g^{pq}R_{\m }^{\ c}R_{\n c}+\frac{3}{2}g_{\m \n }R^{ep}R_{e}^{\ q}-3g_{\n }%
^{\ p}R^{qc}R_{\m c})
\right] \nonumber \\
&& -\frac{1}{2} \Big[ R_{\m \n }R^{cd}R_{cd}+2RR^{cd}R_{\m c\n d} \nonumber \\
&& ~~~ +\nabla _{p}\nabla_{q}(g_{\m \n }g^{pq}R^{cd}R_{cd}+g^{pq}RR_{\m \n }-g_{\m }^{\ p}g_{\n }^{\ q}R^{cd}%
R_{cd}+g_{\m \n }RR^{pq} -2g_{\n }^{\ p}RR_{\m }^{\ q})\Big] \nonumber \\
&&+\frac{1}{12}\left[ 3R^{2}R_{\m \n }+3\nabla _{p}\nabla _{q}%
(g_{\m \n }g^{pq}R^{2}-g_{\m }^{\ p}g_{\n }^{\ q}R^{2})
\right] -\frac{1}{2}g_{\m \n } {\cal L}_5
\eea
and $T_{\m\n}$ is the energy-momentum tensor.

In the absence of any matter sources, the gravity equations of motion have an asymptotically AdS black hole solution
\bea
ds^2 = -f(r) N(r)^2 dt^2 + \frac{dr^2}{f(r)} + \frac{r^2}{L^2} (dx^2 + dy^2 + dz^2)
\eea
When evaluated on these spacetimes, the quasi-topological term reduces to a first order contribution to the gravity equations of motion. Indeed, inserting this metric ansatz into the equations of motion, one obtains that $g(r)=f(r) L^2/r^2$ satisfies
\bea
1 - g(r) + \a g(r)^2 + \b g(r)^3 = \frac{r_H^4}{r^4}
\label{quasitop}
\eea
where $r_H$ is an integration constant which can be interpreted as the horizon of the black hole, because $g(r_H)=0$ is a solution to (\ref{quasitop}). The algebraic equation (\ref{quasitop}) is, of course, exactly solvable. The equation for $N$ is simply $N'(r)=0$, so one chooses $N(r)^2=1/g_\infty$ with
\bea
1 - g_\infty + \a g_\infty^2 + \b g_\infty^3 =0
\eea
to set the dual field theory speed of light to unity \cite{Myers:2010ru}. Note that $N L$ sets the effective AdS radius which is always smaller than $L$.

By direct inspection of the solution to (\ref{quasitop}), the values of $\a$ and $\b$ are constrained by the request for the AdS black hole solution to be stable. The main request is that $f(r\neq \rh)$ is a real, positive definite function. We only consider the range where these constraints are satisfied.

The Hawking temperature of the black hole is given by
\bea
T=\frac{N r_H}{\pi L^2}
\label{temp}
\eea
and can be found, for instance, by using the standard approach of euclidean continuation near the black hole horizon $r_H$. As usual, we interpret (\ref{temp}) as the temperature of the dual CFT on the boundary.

We build our model of holographic superconductivity by adding the matter sources and considering the plane-symmetric metric ansatz
\bea
ds^2 = -f(r) e^{2 \n(r)} dt^2 + \frac{dr^2}{f(r)} + \frac{r^2}{L^2} (dx^2 + dy^2 + dz^2)
\label{ansatzmetric}
\eea
together with the static ansatz for the matter fields
\bea
\begin{split}
A_t &= h(r) \quad \quad A_i = 0, \quad i=t,x,y,z\\
\phi &= \phi(r)
\end{split}
\label{ansatzmatter}
\eea
The scalar field $\phi$ can be chosen to be real without loss of generality.

The normal phase of our holographic superconductor is defined by the solution in which the function $\n(r)$ and the scalar field (also called the scalar hair) vanish identically. The solution is a bit cumbersome, so we only quote here the solution for the gauge field
\bea
h(r) = \r \left( \frac{1}{\rh^2} - \frac{1}{r^2} \right)
\eea
where $\r$ is proportional to the black hole charge, and the Gauss-Bonnet limit of the metric
\bea
f(r) |_{\b \to 0} = \frac{r^2}{2 \a L^2} \left[ 1- \sqrt{1- 4\a \left( 1 - \frac{\rh^4}{r^4} \right) + \frac{8 \a L^2 \k^2 \r^2}{3 r^4 \rh^2} \left( 1 - \frac{\rh^2}{r^2} \right)} \right]
\eea
In the Einstein limit $\a\to 0$ the formula above reduces to the Reissner-Nordstr{\"o}m AdS black hole
\bea
f(r)|_{\a,\b\to 0} = \frac{r^2}{L^2} \left(1-\frac{\rh^4}{r^4} \right) + \frac{2 \k^2 \r^2}{3 r^4} \left(1-\frac{r^2}{\rh^2} \right)
\eea

The superconducting phase corresponds to the hairy black hole solution, in which the scalar field assumes a nontrivial profile. When the temperature is lower than a certain critical value, the charged AdS black hole solution discussed above produces an instability because the scalar effective mass violates the BF bound. Thus, at the critical value for the temperature a phase transition occurs.

Plugging the ansatz (\ref{ansatzmetric}) and (\ref{ansatzmatter}) for the metric and the matter fields into (\ref{gravEOM}), we obtain the following system of equations
\bea
&&\phi'' + \left(\frac{3}{r}+\frac{f'}{f}+\nu'\right) \phi'+ \left(-\frac{m^2}{f}+\frac{q^2 h^2}{e^{2 \nu } f^2}\right) \phi=0 \label{EOM1} \\
&&h'' + \left(\frac{3}{r}-\nu '\right) h' - \frac{2 q^2 \phi ^2}{f} h =0 \label{EOM2} \\
&&\left(1-\frac{2 L^2 \a f}{r^2}-\frac{3 L^4 \b f^2}{r^4}\right) \n '-\frac{2}{3} r \k^2 \left(\frac{q^2 h^2 \phi ^2}{e^{2 \nu } f^2}+\phi'^2\right) =0 \label{EOM3} \\
&&\left(1-\frac{2 L^2 \a f}{r^2}-\frac{3 L^4 \b f^2}{r^4}\right) f' + \frac{2}{r}f - \frac{4 r}{L^2} \nonumber \\ &&~~~~~~~~~~~~~+ \frac{2\k^2 r}{3} \left( \frac{h'^2}{2 e^{2 \n}}+m^2 \phi^2+f \phi'^2+\frac{q^2 h^2 \phi^2}{e^{2 \n} f} \right) + \frac{2 L^4 \b f^3}{r^5} =0
\label{EOM4}
\eea
where a prime indicates a derivative with respect to $r$. These equations and the bulk Lagrangian (\ref{bulkaction}) have different scaling symmetries, already noted in \cite{Hartnoll:2008kx,Barclay:2010up}
\bea
&&r \to ar, t, x^i \to at, ax^i, L \to aL, q \to q/a, m \to m/a; \nonumber \\
&&r \to br, t \to t/b, x^i \to x^i/b, f \to b^2f, \phi \to b\phi; \nonumber \\
&&\phi \to c \phi, \psi \to c\psi, q \to q/c, \kappa^2 \to \kappa^2/c \nonumber
\eea
which we use to set $L=q=\r=1$ in performing numerics. Note that we choose to set $q=1$ as \cite{Barclay:2010up}, so that the probe limit corresponds to $\k\to 0$ instead of $q\to \infty$, and $\r=1$ ensures that the charge parameter is fixed during all our computations.

The Hawking temperature is calculated by
\bea
T=\frac{1}{4\pi} f'(\rh) e^{\n(\rh)}
\eea
In order to interpret it as the boundary field theory temperature, we have to find a solution with the boundary condition
\bea
\n(r) \to 0 \quad {\rm as} \quad r \to \infty
\eea
When we perform the numerics, we use the scaling symmetry \cite{Hartnoll:2008kx}
\bea
e^{2 \n(r)} \to d^2 e^{2 \n(r)}, \qquad \qquad h(r) \to d\, h(r)
\eea
to set to zero the value of the $\n$ function at infinity.

The boundary conditions at the horizon for the gauge and the scalar fields are given by the request of regularity of the equations of motion at the horizon. They read
\bea
h(\rh) = 0 \qquad \qquad
\phi(\rh) = \frac{f'(\rh)}{m^2} \, \phi'(\rh)
\eea
while the metric fields have to obey
\bea
&&
\nu^\prime(\rh)=\frac{2\kappa^2}{3}\rh\left(
\psi^\prime(\rh)^2
+\frac{\phi^\prime(\rh)^2\psi(\rh)^2}{f^\prime(\rh)^2e^{2\nu(\rh)}}
\right)\\
&&
f^\prime(\rh)=\frac{4}{L^2}\rh
-\frac{2\kappa^2}{3}\rh\left(
\frac{\phi^\prime(\rh)^2}{2e^{2\nu(\rh)}}
+m^2\psi(\rh)^2
\right)
\eea
The general asymptotic behavior of the fields near the AdS boundary gives us the field theory data. In particular,
\bea
h(r) = \m - \frac{\r}{r^2} + \ldots \qquad \qquad
\phi(r) = \frac{\phi^{(-)}}{r^{\l_-}} + \frac{\phi^{(+)}}{r^{\l_+}} + \ldots
\eea
where $\m$ and $\r$ are the chemical potential and the charge density, respectively, and $\l_\pm=2\pm \sqrt{4+m^2 L_{eff}^2}$. Here, $L_{eff}= \lim_{r\to\infty} f(r)/r^2$ plays the role of the effective AdS radius; once again, in the limit $\b \to 0$ we recover the Gauss-Bonnet result $L_{eff}^2 = 2 \a L^2 (1-\sqrt{1- 4 \a})^{-1}$.

In order to have a normalizable solution we set
\bea
\phi^{(-)}=0
\label{scalarbnd}
\eea
and read the other quantities from the asymptotic falloffs of the fields. According to the AdS/CFT correspondence, $\phi^{(+)}$ is interpreted as the expectation value $\la {\cal O}_{\l_+} \ra$ of the dual field theory operator with conformal dimension $\l_+$. Note that for a range of masses we could also choose the boundary condition $\phi^{(+)}=0$, but the qualitative features of our results would not be modified \cite{Hartnoll:2008vx,Hartnoll:2008kx}.

For concreteness, we choose $m^2 = -3/L_{eff}^2$. In this way, the dimension of the boundary operator which condenses remains fixed at $\l_+=3$ as we vary the gravity couplings $\a$ and $\b$. The value of the condensate is then read from the asymptotic $r^{-3}$ falloff of the scalar field, when a nontrivial solution exists.

\section{Numerical results for the superconducting phase} \label{numerics}
The behavior of the holographic superconductor is found by integrating the equations of motion (\ref{EOM1})-(\ref{EOM4}) numerically from the horizon to the boundary. We use a shooting method to find the solutions with the appropriate boundary conditions; then, the value of the scalar condensate is obtained by fitting the asymptotic solution. We use this procedure several times with various values of the coupling parameters $\a$ and $\b$ to study how the system reacts to them. Moreover, we also change the backreaction parameter $\k$.

\begin{figure}
\centering
\subfigure{\includegraphics[width=.48\textwidth]{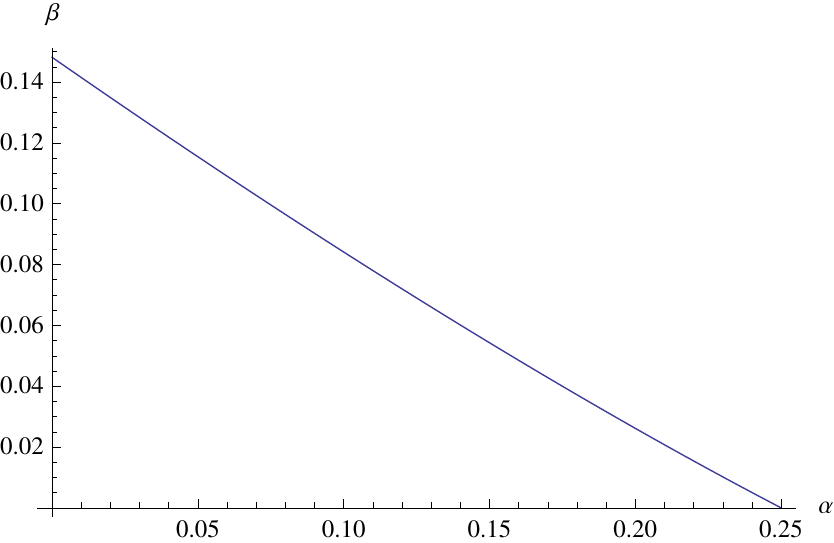}}
\subfigure{\includegraphics[width=.48\textwidth]{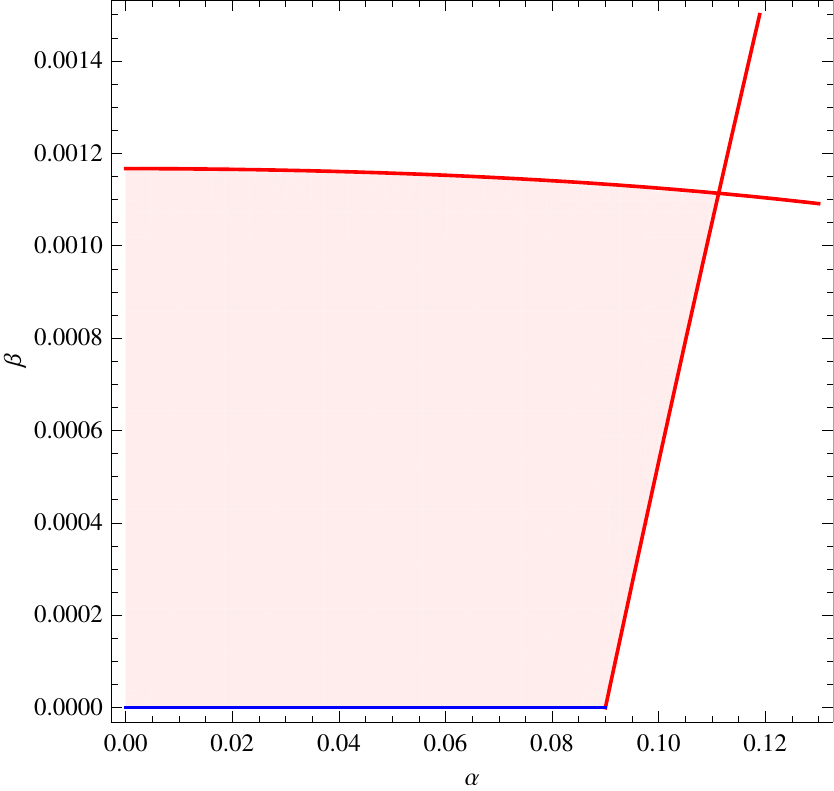}}
\caption{On the left, the line represents the maximum values of $\a$ and $\b$ for which the black hole solution is stable. On the right, the pink region is the allowed parameter range (for positive couplings) as dictated by consistency of the dual CFT. The (blue) bottom line corresponds to five-dimensional GB gravity.}
\label{range}
\end{figure}
The values of $\a$ and $\b$ allowed by the condition to obtain a stable solution are shown in left plot of figure \ref{range}. However, requiring that the AdS solution has a consistent dual field theory poses severe bounds on the couplings. In particular, it has been shown that the requirements of causality and positivity of energy fluxes in the CFT restrict the allowed parameter region as shown in the right plot of figure \ref{range} \cite{Myers:2010ru}. Thus, we only consider this smaller parameter region.

It has been shown \cite{Hartnoll:2008kx,Barclay:2010up} that the effect of the backreaction is to decrease the critical temperature, making the condensation harder. The authors explained this phenomenon by noticing that when the backreaction parameter is increased the scalar field not only screens the charge of the black hole, but also its mass: thus, for a given charge and temperature, the radius of the black hole is increased, which makes harder for the scalar field itself to condense.

Another interesting feature shown in \cite{Barclay:2010up} is that as one increases $\a$, the critical temperature decreases until it reaches a turning point and starts to increase again. In particular, the higher is the backreaction, the more pronounced is the minimum, and the critical temperature in the limit $\a=1/4$ can be even higher than that in the Einstein limit $\a=0$. However, we note that all the turning points found in \cite{Barclay:2010up} are outside the allowed parameter space shown in the right plot of figure \ref{range}.

\begin{figure}
\center
\includegraphics{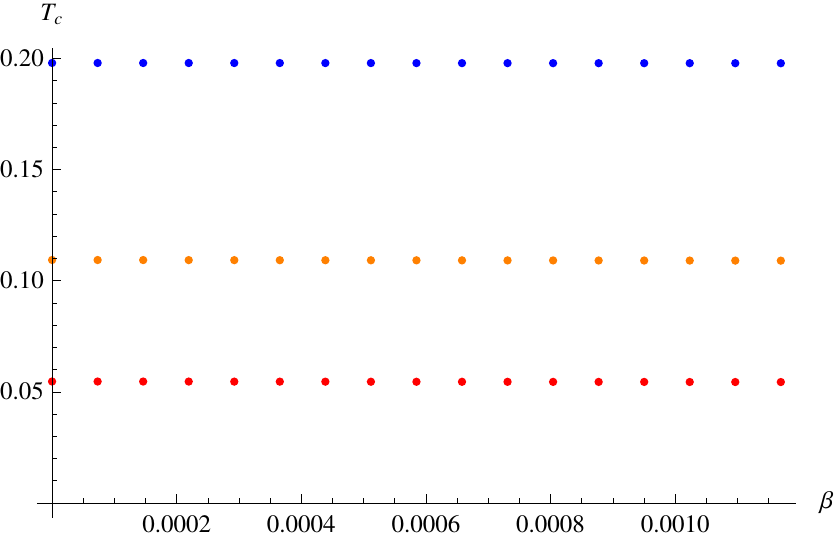}
\caption{The critical temperature as a function of the curvature-cubed parameter $\b$ at $\a=0$. The top (blue) data are for $\k^2=0$ (i.e. no backreaction), the orange points for $\k^2=0.1$, while the bottom (red) ones refer to $\k^2=0.2$.}
\label{tc1}
\end{figure}
We first study how the critical temperature changes as one explores the range of the allowed parameters. We particularly concentrate on the allowed region for the curvature-cubed parameter. A plot of the critical temperature as a function of $\b$ is given in figure \ref{tc1} for three values of the backreaction parameter: the top line represents the non-backreacting case, the mid one is computed with a backreaction parameter $\k^2=0.1$ while $\k^2=0.2$ for the bottom line. It is evident that increasing the backreaction parameter lowers the critical temperature.

\begin{figure}
\centering
\subfigure{\includegraphics[width=.47\textwidth]{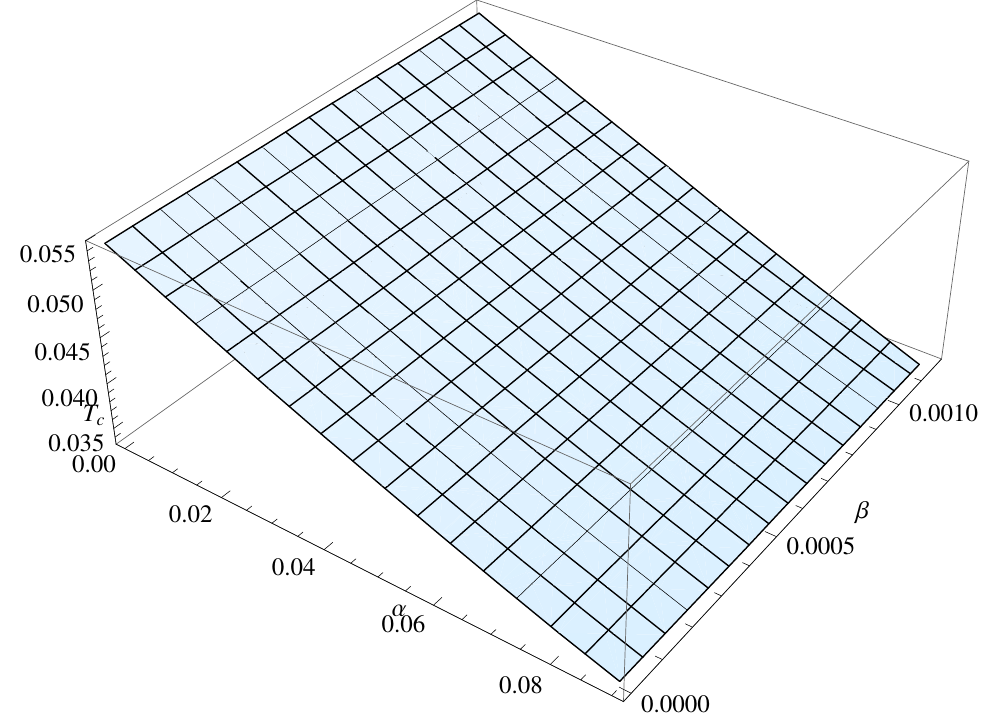}}
\hspace{2mm}
\subfigure{\includegraphics[width=.47\textwidth]{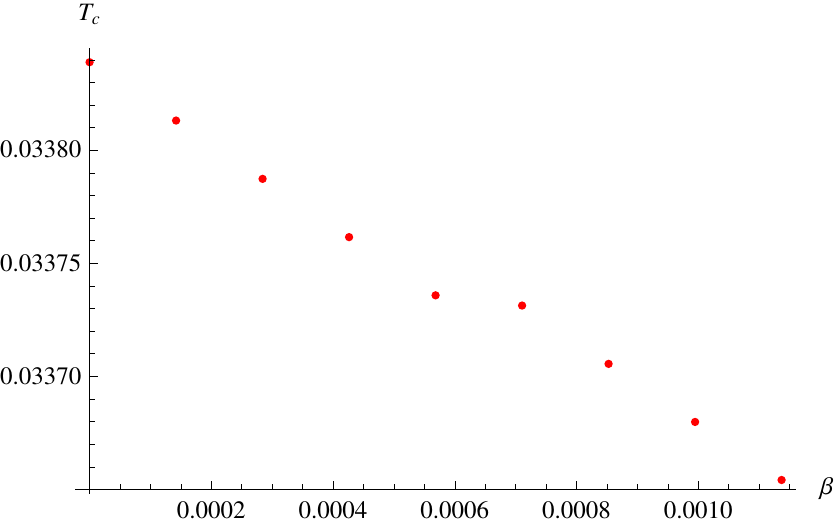}}
\caption{On the left, the critical temperature as a function of the Gauss-Bonnet and the curvature-cubed parameter $\a$ and $\b$. On the right, the $\a=9/100$ slice of the left plot. The backreaction parameter is fixed at $\k^2=0.2$.}
\label{tc2}
\end{figure}
The critical temperature is also affected by the gravitational couplings $\a$ and $\b$. In particular, we push our analysis towards the highest curvature-cubed parameter admitted by consistency of the dual CFT. From our numerical results, we see that the effect of the curvature-cubed term results in a lowering of the critical temperature, without any appearance of turning points, thus making the condensation harder. However, the lowering of the critical temperature is of order of one part per hundred in the allowed parameter range and it is not evident in figure \ref{tc1}. A plot of the critical temperature at fixed $\k^2=0.2$ as a function of the two gravity coupling parameters is represented in figure \ref{tc2}. We note that the effects of the Gauss-Bonnet term are much more evident, as its allowed range is larger. The right plot of figure \ref{tc2} shows that the critical temperature is a monotonically decreasing function of $\b$ at fixed $\a=9/100$. A similar behavior also holds for the other values of $\a$. The qualitative behavior of figure \ref{tc2} is in agreement with the results presented in \cite{Kuang:2010jc}, where the non-backreacting case of the quasi-topological gravity is considered.

\begin{figure}
\center
\includegraphics{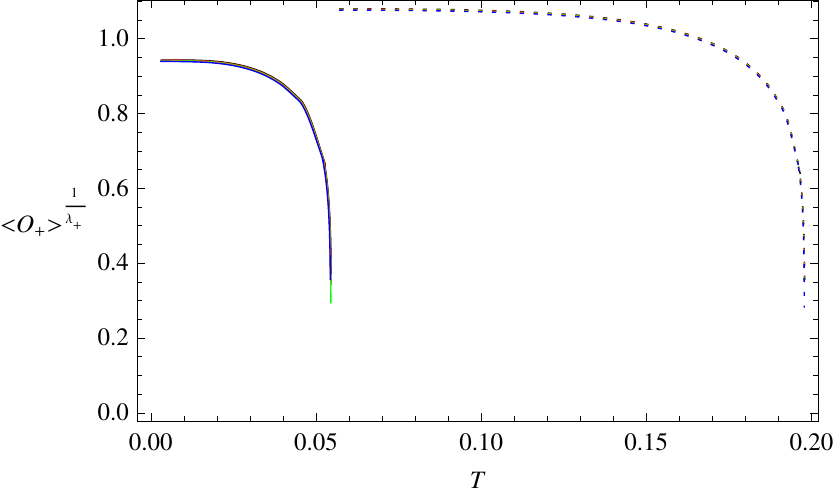}
\caption{The scalar condensate as a function of the temperature. The Gauss-Bonnet parameter is kept fixed at zero. There are five almost coincident lines, corresponding to values of $\b$ which run from 0 to 0.00117, with equally spaced intervals of 0.00029. The undulating lines are a numerical artifact. The dotted lines have been obtained for a backreaction parameter $\k=0$, while the solid ones for $\k^2=0.2$.}
\label{condensate1}
\end{figure}

Figure \ref{condensate1}
\begin{figure}
\center
\includegraphics{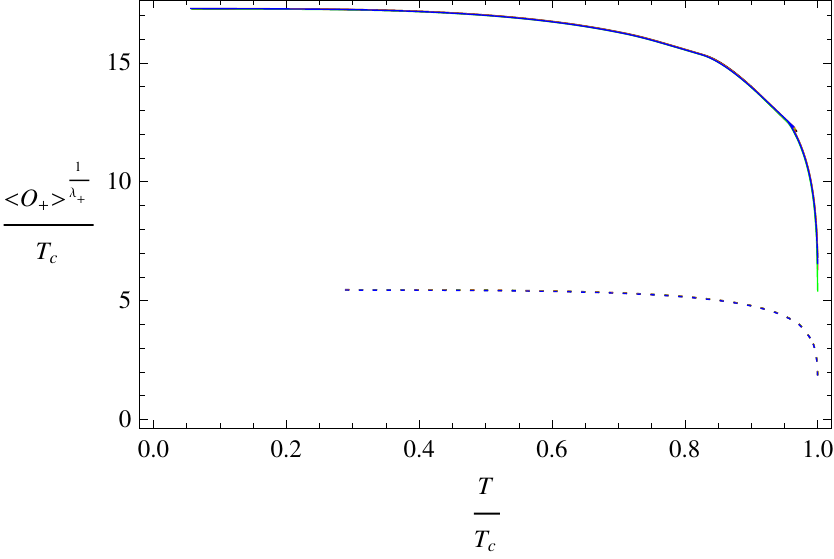}
\caption{The scalar condensate as a function of the temperature normalized with the critical temperature. The Gauss-Bonnet parameter is kept fixed at zero, while the lines correspond to values of $\b$ which run from 0 to 0.00117, with equally spaced intervals of 0.00029. The undulating lines are a numerical artifact. The dotted lines have been obtained for a backreaction parameter $\k=0$, while the solid ones for $\k^2=0.2$.}
\label{condensate2}
\end{figure}
shows the values of the scalar condensate as a function of the temperature for various values of $\k^2$ and $\b$, keeping $\a$ fixed. From this figure it is evident that the backreaction lowers both the critical temperature and the value of the condensate, thus making the condensation harder. Due to the bounds on the higher curvature couplings, the variation of the scalar condensate with respect to $\b$ is not appreciable. We numerically find that it monotonically decreases. Figure \ref{condensate2} shows the same curves normalized by the critical temperature, so that we deal with a dimensionless function of a dimensionless variable. Because the critical temperature is lowered when either the backreaction or curvature-cubed parameter are enhanced, the height of the dimensionless condensate is higher. This happens despite the fact the unnormalized curves tend to decrease. Note that the effect of the backreaction is very evident, while the effect of the quasi-topological parameter is negligible.

\section{Conductivity} \label{sec:conductivity}
According to the AdS/CFT correspondence, the bulk field $A_\m$ corresponds to a four-current $J_\m$ on the boundary. To compute the conductivity of our system, we have to introduce a perturbation of the Maxwell field $A_\m$ on top of the hairy black hole solution. Writing the perturbation as $A(t,r)=A_x(r) e^{-i \omega t} dx$ we obtain the linearized equation of motion for $A_x$:
\bea
A^{\prime\prime}+\left(\frac{f^\prime}{f}+\nu^\prime+\frac{1}{r}\right)A^\prime
+\left[\frac{\omega^2}{f^2e^{2\nu}}-\frac{2}{f}q^2\psi^2
-\frac{2\kappa^2r^2\phi^{\prime2}}{fe^{2\nu}\left(r^2-2\alpha f-3 \b f^3/r^2\right)}
\right]A =0 \;.
\label{A:eq}
\eea
where we set $L=1$ for simplicity. The equation has to be solved under the incoming boundary condition near the horizon
\bea
A(r) \sim f(r)^{-i \frac{\omega}{4 \pi T}}
\eea
where $T$ is the Hawking temperature of the black hole. In the asymptotic AdS region ($r \to \infty$) the general solution behaves as
\bea
A=a_0 + \frac{a_2}{r^2}
+\frac{a_0 L_{eff}^4 \omega^2}{2r^2}
\log r
\eea
where $a_0$ and $a_2$ are integration constants. The logarithmic term suffers from an ambiguity scale, related to the fact that it leads to a divergence in the Green function that has to be removed by an appropriate counterterm. We use here the same conventions as \cite{Barclay:2010up}.

Given the results of the previous section, we numerically solve the equation for the Maxwell field perturbation with the given boundary condition and obtain its large $r$ coefficients. Once the asymptotic parameters have been computed, the conductivity is given by the formula
\bea
\s = \frac{2 a_2}{i \omega L_{eff}^4 a_0} + \frac{i \omega}{2} - i \omega \log L_{eff}
\eea
where the arbitrariness of scale reflects its presence in the linear term in $\omega$.

\begin{figure}
\center
\includegraphics{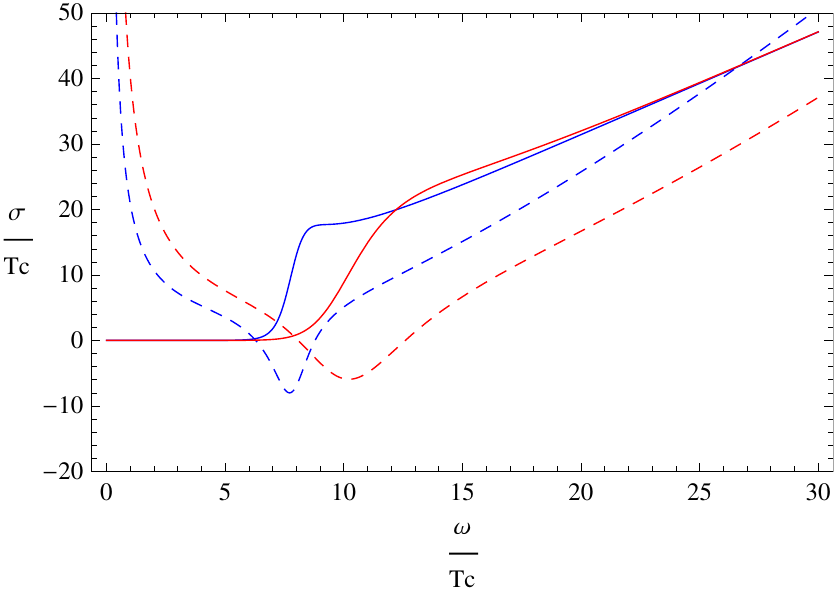}
\subfigure{\includegraphics[width=.48\textwidth]{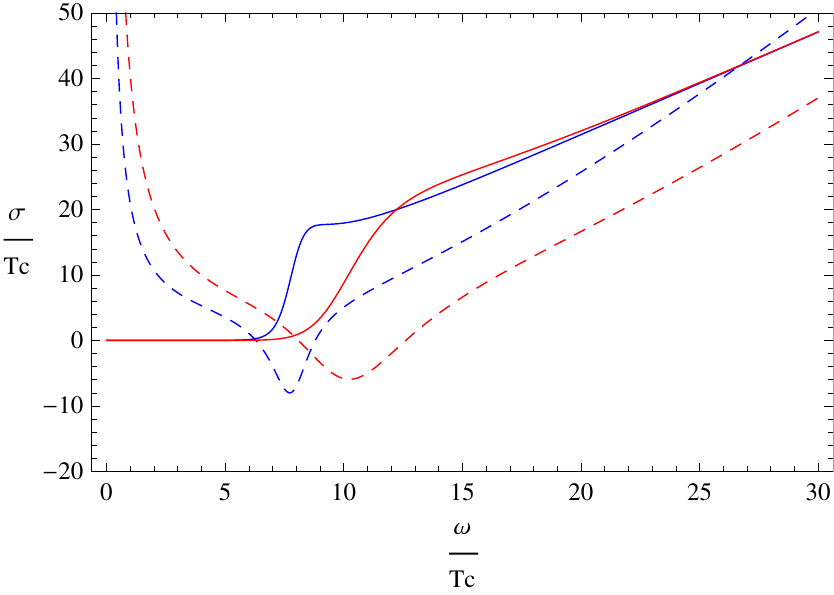}} \quad
\subfigure{\includegraphics[width=.48\textwidth]{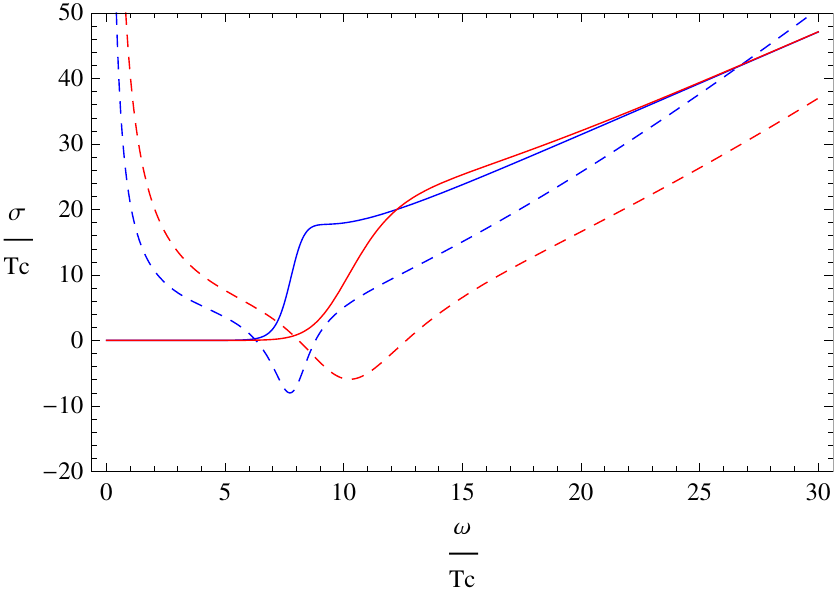}}
\caption{Plots of the real (solid lines) and imaginary (dotted lines) parts of the conductivity as a function of the frequency, normalized with the critical temperature. The non-backreacting case is shown in blue, while the backreacted $\k^2=0.05$ case is shown in red. All the plots have $\a=0$. The upper plot has $\b=0$, the lower left one $\b=0.00058$ and the lower right one $\b=0.00117$.}
\label{conductivity}
\end{figure}
In figure \ref{conductivity} we plot our numerical results for the electrical conductivity of the system. Once again we normalize both the conductivity and the frequency with the critical temperature. In all cases, the solid line represents the real part of the conductivity, while the dotted line represents the imaginary part. The conductivity for the non-backreacted case ($\k^2=0$) is shown in blue, that for the backreacted case (we choose $\k^2=0.05$) in red. The qualitative features do not change as we vary $\k^2$. The upper figure has $\a=\b=0$, the lower ones have $\a=0, \b=0.0006$ (left) and  $\a=0, \b=0.0011$ (right).

It has been observed \cite{Horowitz:2008bn} that in field theories with Einstein dual the real part of the conductivity has a gap with frequency $\omega_g$, where the relation
\bea
\frac{\omega_g}{T_c} \simeq 8
\label{freqgap}
\eea
holds. Here, $\omega_g$ is the value for the frequency at which the imaginary part of the conductivity envelops a global minimum. In the case of field theories with Einstein-Gauss-Bonnet gravity dual this is no longer true, and indeed the normalized frequency gap highly depends on the values of the backreaction and Gauss-Bonnet parameters. In particular, increasing $\a$ pushes the frequency gap to higher values without affecting the critical temperature very much, while the frequency gap does not change but the critical temperature decreases when the backreaction is switched on. It should be noted, however, that in the CFT dual allowed parameter range (right plot of figure \ref{range}) the effects of the Gauss-Bonnet term consist in a deviation from (\ref{freqgap}) of order of one percent. From our results in figure \ref{conductivity}, we see that the same is true also when the quasi-topological term is considered. Indeed, changing the backreaction parameter results in a lowering of the critical temperature, while increasing the $\b$ parameter does not alter the critical temperature very much but increases the frequency gap. Moreover, the backreaction smooths both the step in the real part of the conductivity and the dip in the imaginary part. The plots in figure \ref{conductivity} looks almost identical because the allowed range of $\b$ is very restricted.

\section{Conclusions} \label{conclusions}
In this paper we studied the effects of the backreaction on the quasi-topological holographic superconductors. The non-backreacting case was studied in \cite{Kuang:2010jc} and our results in the probe limit agree with those presented there. We only considered the simplest model with a quadratic potential term for the scalar field which represents the gravity dual of the boundary operator which condenses. For a wide range of parameters, included the backreaction parameter, we numerically studied the condensation of the scalar field as a function of the temperature. We found that as the bulk parameters are increased, the value of the scalar condensate decreases. Moreover, the critical temperature of the dual field theory also decreases, and this is more evident when the backreaction is taken into account. Thus, both the higher derivative corrections and the backreaction make the condensation harder. Finally, we found that the value of the scalar condensate normalized with the critical temperature is higher. Our results complete and generalize those found in \cite{Barclay:2010up} and \cite{Kuang:2010jc}. In the former, a more careful treatment of the Gauss-Bonnet coupling has been carried out, and it reveals more subtle than the curvature-cubed corrections presented here.

We numerically calculated the electrical conductivity of the holographic superconductor. We found that the so-called universal relation $\omega_g\simeq 8 T_c$ is spoiled by the effects of the backreaction but it is not affected very much by the quasi-topological term in the bulk action when the bounds coming from the consistency of the dual CFT are taken into account. The backreaction does not affect $\omega_g$ but decreases the temperature, while the opposite is true for the curvature-cubed parameter. In all cases, $\omega_g/T_c$ is an increasing function of the parameters we considered.

In addition to higher derivative terms to the gravitational part of the action one could also consider adding corrections to the matter part of the action. The latter are typical in string-derived models, and they should be taken into account if one wants to describe the low energy dynamics of a general string background. While field theories with a dual higher derivative matter Lagrangian have been studied \cite{Myers:2009ij}, it is interesting to study their effects in the context of holographic superconductivity \cite{Siani:2010??}.

We did not derive the gravitational theory from a consistent truncation of string theory. Thus, we only know that in the dual field theory a charged scalar operator condenses, but we do not know the microscopic details of the dual field theory. Moreover, a consistent truncation of string theory can uniquely fix the scalar potential to a more complicated form than that studied here. In this sense, it would be interesting to study more general potentials as well as nonabelian extensions which are very common in string theory constructions.

\section*{Acknowledgments}

We thank Silvia Penati for encouragements. We also thank Silvia Penati and Alessandro Tomasiello for useful discussions and for reading the manuscript.

This work is supported in part by INFN and PRIN under contract
prot.2007-5ATT78-002.


\begin{thebibliography}{99}


%\cite{Maldacena:1997re}
\bibitem{Maldacena:1997re}
  J.~M.~Maldacena,
  %``The large N limit of superconformal field theories and supergravity,''
  Adv.\ Theor.\ Math.\ Phys.\  {\bf 2} (1998) 231
  [Int.\ J.\ Theor.\ Phys.\  {\bf 38} (1999) 1113]
  [arXiv:hep-th/9711200].
  %%CITATION = IJTPB,38,1113;%%

%\cite{Hartnoll:2009sz}
\bibitem{Hartnoll:2009sz}
  S.~A.~Hartnoll,
  %``Lectures on holographic methods for condensed matter physics,''
  Class.\ Quant.\ Grav.\  {\bf 26} (2009) 224002
  [arXiv:0903.3246 [hep-th]].
  %%CITATION = CQGRD,26,224002;%%

%\cite{Herzog:2009xv}
\bibitem{Herzog:2009xv}
  C.~P.~Herzog,
  %``Lectures on Holographic Superfluidity and Superconductivity,''
  J.\ Phys.\ A  {\bf 42} (2009) 343001
  [arXiv:0904.1975 [hep-th]].
  %%CITATION = JPAGB,A42,343001;%%

%\cite{Gubser:2005ih}
\bibitem{Gubser:2005ih}
  S.~S.~Gubser,
  %``Phase transitions near black hole horizons,''
  Class.\ Quant.\ Grav.\  {\bf 22} (2005) 5121
  [arXiv:hep-th/0505189].
  %%CITATION = CQGRD,22,5121;%%

%\cite{Gubser:2008px}
\bibitem{Gubser:2008px}
  S.~S.~Gubser,
  %``Breaking an Abelian gauge symmetry near a black hole horizon,''
  Phys.\ Rev.\  D {\bf 78} (2008) 065034
  [arXiv:0801.2977 [hep-th]].
  %%CITATION = PHRVA,D78,065034;%%

%\cite{Hartnoll:2008vx}
\bibitem{Hartnoll:2008vx}
  S.~A.~Hartnoll, C.~P.~Herzog and G.~T.~Horowitz,
  %``Building a Holographic Superconductor,''
  Phys.\ Rev.\ Lett.\  {\bf 101} (2008) 031601
  [arXiv:0803.3295 [hep-th]].
  %%CITATION = PRLTA,101,031601;%%

%\cite{Hartnoll:2008kx}
\bibitem{Hartnoll:2008kx}
  S.~A.~Hartnoll, C.~P.~Herzog and G.~T.~Horowitz,
  %``Holographic Superconductors,''
  JHEP {\bf 0812} (2008) 015
  [arXiv:0810.1563 [hep-th]].
  %%CITATION = JHEPA,0812,015;%%

%\cite{Horowitz:2008bn}
\bibitem{Horowitz:2008bn}
  G.~T.~Horowitz and M.~M.~Roberts,
  %``Holographic Superconductors with Various Condensates,''
  Phys.\ Rev.\  D {\bf 78}, 126008 (2008)
  [arXiv:0810.1077 [hep-th]].
  %%CITATION = PHRVA,D78,126008;%%

%\cite{Gubser:2009cg}
\bibitem{Gubser:2009cg}
  S.~S.~Gubser and A.~Nellore,
  %``Ground states of holographic superconductors,''
  Phys.\ Rev.\  D {\bf 80}, 105007 (2009)
  [arXiv:0908.1972 [hep-th]].
  %%CITATION = PHRVA,D80,105007;%%

%\cite{Horowitz:2009ij}
\bibitem{Horowitz:2009ij}
  G.~T.~Horowitz and M.~M.~Roberts,
  %``Zero Temperature Limit of Holographic Superconductors,''
  JHEP {\bf 0911}, 015 (2009)
  [arXiv:0908.3677 [hep-th]].
  %%CITATION = JHEPA,0911,015;%%

%\cite{Konoplya:2009hv}
\bibitem{Konoplya:2009hv}
  R.~A.~Konoplya and A.~Zhidenko,
  %``Holographic conductivity of zero temperature superconductors,''
  Phys.\ Lett.\  B {\bf 686}, 199 (2010)
  [arXiv:0909.2138 [hep-th]].
  %%CITATION = PHLTA,B686,199;%%

%\cite{Gubser:2008wv}
\bibitem{Gubser:2008wv}
  S.~S.~Gubser and S.~S.~Pufu,
  %``The gravity dual of a p-wave superconductor,''
  JHEP {\bf 0811}, 033 (2008)
  [arXiv:0805.2960 [hep-th]].
  %%CITATION = JHEPA,0811,033;%%

%\cite{Peeters:2009sr}
\bibitem{Peeters:2009sr}
  K.~Peeters, J.~Powell and M.~Zamaklar,
  %``Exploring colourful holographic superconductors,''
  JHEP {\bf 0909}, 101 (2009)
  [arXiv:0907.1508 [hep-th]].
  %%CITATION = JHEPA,0909,101;%%

%\cite{Maeda:2009wv}
\bibitem{Maeda:2009wv}
  K.~Maeda, M.~Natsuume and T.~Okamura,
  %``Universality class of holographic superconductors,''
  Phys.\ Rev.\  D {\bf 79}, 126004 (2009)
  [arXiv:0904.1914 [hep-th]].
  %%CITATION = PHRVA,D79,126004;%%

%\cite{Nakano:2008xc}
\bibitem{Nakano:2008xc}
  E.~Nakano and W.~Y.~Wen,
  %``Critical Magnetic Field In A Holographic Superconductor,''
  Phys.\ Rev.\  D {\bf 78}, 046004 (2008)
  [arXiv:0804.3180 [hep-th]].
  %%CITATION = PHRVA,D78,046004;%%

%\cite{Albash:2008eh}
\bibitem{Albash:2008eh}
  T.~Albash and C.~V.~Johnson,
  %``A Holographic Superconductor in an External Magnetic Field,''
  JHEP {\bf 0809}, 121 (2008)
  [arXiv:0804.3466 [hep-th]].
  %%CITATION = JHEPA,0809,121;%%

%\cite{Ge:2010aa}
\bibitem{Ge:2010aa}
  X.~-H.~Ge, B.~Wang, S.~-F.~Wu {\it et al.},
  %``Analytical study on holographic superconductors in external magnetic field,''
  JHEP {\bf 1008 } (2010)  108.
  [arXiv:1002.4901 [hep-th]].

%\cite{Takahashi:2009dz}
\bibitem{Takahashi:2009dz}
  T.~Takahashi and J.~Soda,
  %``Stability of Lovelock Black Holes under Tensor Perturbations,''
  Phys.\ Rev.\  D {\bf 79}, 104025 (2009)
  [arXiv:0902.2921 [gr-qc]].
  %%CITATION = PHRVA,D79,104025;%%

%\cite{Kanno:2010pq}
\bibitem{Kanno:2010pq}
  S.~Kanno and J.~Soda,
  %``Stability of Holographic Superconductors,''
  arXiv:1007.5002 [hep-th].
  %%CITATION = ARXIV:1007.5002;%%

%\cite{Brihaye:2010mr}
\bibitem{Brihaye:2010mr}
  Y.~Brihaye, B.~Hartmann,
  %``Holographic Superconductors in 3+1 dimensions away from the probe limit,''
  Phys.\ Rev.\  {\bf D81}, 126008 (2010).
  [arXiv:1003.5130 [hep-th]].

%\cite{Gubser:2009qm}
\bibitem{Gubser:2009qm}
  S.~S.~Gubser, C.~P.~Herzog, S.~S.~Pufu and T.~Tesileanu,
  %``Superconductors from Superstrings,''
  Phys.\ Rev.\ Lett.\  {\bf 103}, 141601 (2009)
  [arXiv:0907.3510 [hep-th]].
  %%CITATION = PRLTA,103,141601;%%

%\cite{Gauntlett:2009dn}
\bibitem{Gauntlett:2009dn}
  J.~P.~Gauntlett, J.~Sonner and T.~Wiseman,
  %``Holographic superconductivity in M-Theory,''
  Phys.\ Rev.\ Lett.\  {\bf 103} (2009) 151601
  [arXiv:0907.3796 [hep-th]].
  %%CITATION = PRLTA,103,151601;%%

%\cite{Gregory:2009fj}
\bibitem{Gregory:2009fj}
  R.~Gregory, S.~Kanno and J.~Soda,
  %``Holographic Superconductors with Higher Curvature Corrections,''
  JHEP {\bf 0910}, 010 (2009)
  [arXiv:0907.3203 [hep-th]].
  %%CITATION = JHEPA,0910,010;%%

%\cite{Pan:2009xa}
\bibitem{Pan:2009xa}
  Q.~Pan, B.~Wang, E.~Papantonopoulos, J.~Oliveira and A.~B.~Pavan,
  % ``Holographic Superconductors with various condensates in
  %Einstein-Gauss-Bonnet gravity,''
  Phys.\ Rev.\  D {\bf 81}, 106007 (2010)
  [arXiv:0912.2475 [hep-th]].
  %%CITATION = PHRVA,D81,106007;%%

%\cite{Pan:2010at}
\bibitem{Pan:2010at}
  Q.~Pan and B.~Wang,
  %``General holographic superconductor models with Gauss-Bonnet corrections,''
  arXiv:1005.4743 [hep-th].
  %%CITATION = ARXIV:1005.4743;%%

%\cite{Cai:2010cv}
\bibitem{Cai:2010cv}
  R.~G.~Cai, Z.~Y.~Nie and H.~Q.~Zhang,
  %``Holographic p-wave superconductors from Gauss-Bonnet gravity,''
  arXiv:1007.3321 [hep-th].
  %%CITATION = ARXIV:1007.3321;%%

%\cite{Barclay:2010up}
\bibitem{Barclay:2010up}
  L.~Barclay, R.~Gregory, S.~Kanno and P.~Sutcliffe,
  %``Gauss-Bonnet Holographic Superconductors,''
  arXiv:1009.1991 [hep-th].
  %%CITATION = ARXIV:1009.1991;%%

%\cite{Oliva:2010eb}
\bibitem{Oliva:2010eb}
  J.~Oliva and S.~Ray,
  %``A new cubic theory of gravity in five dimensions: Black hole, Birkhoff's
  %theorem and C-function,''
  arXiv:1003.4773 [gr-qc].
  %%CITATION = ARXIV:1003.4773;%%

%\cite{Myers:2010ru}
\bibitem{Myers:2010ru}
  R.~C.~Myers and B.~Robinson,
  %``Black Holes in Quasi-topological Gravity,''
  JHEP {\bf 1008} (2010) 067
  [arXiv:1003.5357 [gr-qc]].
  %%CITATION = JHEPA,1008,067;%%

%\cite{Myers:2010jv}
\bibitem{Myers:2010jv}
  R.~C.~Myers, M.~F.~Paulos and A.~Sinha,
  %``Holographic studies of quasi-topological gravity,''
  JHEP {\bf 1008}, 035 (2010)
  [arXiv:1004.2055 [hep-th]].
  %%CITATION = JHEPA,1008,035;%%

%\cite{Amsel:2010aj}
\bibitem{Amsel:2010aj}
  A.~J.~Amsel and D.~Gorbonos,
  % ``The Weak Gravity Conjecture and the Viscosity Bound with Six-Derivative
  %Corrections,''
  arXiv:1005.4718 [hep-th].
  %%CITATION = ARXIV:1005.4718;%%

%\cite{Myers:2010xs}
\bibitem{Myers:2010xs}
  R.~C.~Myers and A.~Sinha,
  %``Seeing a c-theorem with holography,''
  Phys.\ Rev.\  D {\bf 82} (2010) 046006
  [arXiv:1006.1263 [hep-th]].
  %%CITATION = PHRVA,D82,046006;%%

%\cite{Sinha:2010pm}
\bibitem{Sinha:2010pm}
  A.~Sinha,
  %``On higher derivative gravity, c-theorems and cosmology,''
  arXiv:1008.4315 [hep-th].
  %%CITATION = ARXIV:1008.4315;%%

%\cite{Kuang:2010jc}
\bibitem{Kuang:2010jc}
  X.~M.~Kuang, W.~J.~Li and Y.~Ling,
  %``Holographic Superconductors in Quasi-topological Gravity,''
  arXiv:1008.4066 [hep-th].
  %%CITATION = ARXIV:1008.4066;%%

%\cite{Myers:2009ij}
\bibitem{Myers:2009ij}
  R.~C.~Myers, M.~F.~Paulos, A.~Sinha,
  %``Holographic Hydrodynamics with a Chemical Potential,''
  JHEP {\bf 0906 } (2009)  006.
  [arXiv:0903.2834 [hep-th]].

%\cite{Siani:2010??}
\bibitem{Siani:2010??}
  M.~Siani,
  in preparation.


\end{thebibliography}
\end{document}